\documentclass[conference]{IEEEtran}
\IEEEoverridecommandlockouts
\usepackage{cite}
\usepackage{amsmath,amssymb,amsfonts}
\usepackage{algorithmic}
\usepackage{graphicx}
\usepackage{textcomp}
\ifCLASSOPTIONcompsoc
\usepackage[caption=false,font=normalsize,labelfon
t=sf,textfont=sf]{subfig}
\else
\usepackage[caption=false,font=footnotesize]{subfig}
\fi
\usepackage{subcaption}
\usepackage{xcolor}

\usepackage{listings}
\lstdefinelanguage{yaml}{
  keywords={true,false,null,y,n},
  sensitive=false,
  comment=[l]{\#},
  morestring=[b]',
  morestring=[b]"
}
\def\BibTeX{{\rm B\kern-.05em{\sc i\kern-.025em b}\kern-.08em
    T\kern-.1667em\lower.7ex\hbox{E}\kern-.125emX}}
\begin{document}

\title{SDVDiag: Using Context-Aware Causality Mining for the Diagnosis of Connected Vehicle Functions\\
}

\author{
\IEEEauthorblockN{Matthias Weiß, Falk Dettinger, Elias Detrois, Nasser Jazdi and Michael Weyrich}
\IEEEauthorblockA{
\textit{Institute of Industrial Automation and Software Engineering (IAS)} \\
\textit{University of Stuttgart} \\
Pfaffenwaldring 47, 70550 Stuttgart, Germany \\
E-Mail: \{matthias.weiss, falk.dettinger, elias.detrois, nasser.jazdi, michael.weyrich\}@ias.uni-stuttgart.de}}


\maketitle

\begin{abstract}
Real-world implementations of connected vehicle functions are spreading steadily, yet operating these functions reliably remains challenging due to their distributed nature and the complexity of the underlying cloud, edge, and networking infrastructure. Quick diagnosis of problems and understanding the error chains that lead to failures is essential for reducing downtime. However, diagnosing these systems is still largely performed manually, as automated analysis techniques are predominantly data-driven and struggle with hidden relationships and the integration of context information.

This paper addresses this gap by introducing a multimodal approach that integrates human feedback and system-specific information into the causal analysis process. Reinforcement Learning from Human Feedback is employed to continuously train a causality mining model while incorporating expert knowledge. Additional modules leverage distributed tracing data to prune false-positive causal links and enable the injection of domain-specific relationships to further refine the causal graph.

Evaluation is performed using an automated valet parking application operated in a connected vehicle test field. Results demonstrate a significant increase in precision from 14\% to 100\% for the detection of causal edges and improved system interpretability compared to purely data-driven approaches, highlighting the potential for system operators in the connected vehicle domain.


\end{abstract}


\section{Introduction}

In modern vehicles, functions are becoming increasingly complex and capable \cite{Stuempfle2025SDV, nezami2025computing, dettinger2024survey}, requiring not only local data but also external information and communication with other traffic participants. Ensuring reliable operation of these connected functions remains a major challenge due to their distributed nature across multiple network layers \cite{almuseelem2025deep, dettinger2025directives}. As vehicles communicate with edge nodes and cloud infrastructure through volatile data streams, even minor software issues in such distributed architectures can have significant consequences for overall functionality, often leading to downtime and economic losses \cite{zhang2024latency, liu2024multi}.

Given these circumstances, rapid diagnosis of problems becomes essential, including understanding the chain of errors that led to the respective failure. To address this, data-driven techniques have been proposed for the mining of causal relationships, which represent how system components depend on each other in a given system state. However, these learning-based approaches struggle with hidden dependencies and the integration of human knowledge or other contextual information, ultimately limiting their effectiveness \cite{wang2024comprehensivesurveyrootcause}. This challenge is further compounded by the fact that connected vehicle systems undergo continuous change due to frequent updates and varying network conditions \cite{weiss2023continuous}. Such dynamics necessitate frequent retraining, which is unsuitable for continuous operation. As a result, diagnosing system failures remains largely dependent on manual intervention by human experts, causing outages to persist for extended periods and requiring substantial investment to mitigate damages.

To address these challenges, this paper proposes a novel approach for continuous system analysis that combines automated techniques with human knowledge and integrates contextual information specific to connected vehicle applications. The main contributions are: (1) a reinforcement learning framework that efficiently incorporates human feedback into the causal discovery process while preserving it in a knowledge database, (2) a multimodal analysis pipeline that simultaneously captures different performance metrics and the system's topology, and (3) a rule-based interface for integrating context information, including hidden relationships and application data.

The remainder of this paper is structured as follows. Section~\ref{sec:related_work} reviews related work on causal discovery and reinforcement learning approaches for system diagnosis. Section~\ref{sec:concept} presents the proposed concept, detailing the data collection pipeline, the human feedback integration mechanism, and the context-aware refinement modules. Section~\ref{sec:evaluation} evaluates the approach using an automated valet parking application, demonstrating improvements in causal graph quality and root cause analysis. Section~\ref{sec:conclusion} concludes the paper and outlines directions for future work.


\section{Background}\label{sec:related_work}

The drive to enhance user experience, particularly in comfort functions and assisted or autonomous driving \cite{dettinger2025directives}, has accelerated the adoption of distributed functions \cite{dettinger2024future}. Their introduction entails profound implications for system architectures and diagnosis processes.


\subsection{Connected Vehicle Application Architectures}
Within connected vehicle architectures, distributed functions often rely on sub‑functions that are hosted centrally in the network backend \cite{weiss2025opensource}. This centralization allows all participants with backend access to utilize these functions. To ensure broad availability, functions must therefore be provisioned in a scalable manner, independent of their execution location or context. Such independent software units are referred to as \textit{services} and are typically deployed and replicated across computing clusters. 

In the context of connected vehicles, these services typically require a low latency and high reliability. \textit{Multi‑Access Edge Computing (MEC)} is therefore employed in this context, to extend cloud capabilities to the edge of the network, placing computation and storage resources closer to vehicles and infrastructure \cite{Stuempfle2025SDV}. By reducing the physical and logical distance between data sources and processing units, MEC enables low‑latency and real‑time execution of latency‑sensitive services. Beyond pure computation, MEC environments also provide scalability by allowing services to be dynamically deployed and replicated across edge nodes, ensuring continuous availability even under volatile network conditions. 

In addition, a MEC environment supports a variety of auxiliary functions that extend beyond pure computation. These include function offloading \cite{dettinger2025directives, tao2025bidding}, where vehicles delegate the processing of resource‑intensive tasks to nearby edge servers, as well as diagnostic services that monitor system states \cite{weiss2024review}. Together, these capabilities make MEC a key enabler for the execution of various applications in connected vehicles.



\subsection{Causal Inference in Distributed Systems}\label{sec:causal_inference}

With the increasing complexity and number of components in modern distributed systems, causal chains between events and subsystems become longer and more intricate. Understanding these causal relationships is essential for ensuring reliability, diagnosing failures, and optimizing performance \cite{weiss2025sdvdiag}. Such causal chains can be represented using \textit{Bayesian networks} or \textit{Structural Causal Models}, which encode dependencies between variables and allow reasoning about interventions, hidden confounders, and counterfactual scenarios \cite{hernan2025causal}. Applications of causal inference in distributed systems include root cause analysis, hypothesis testing about system behavior, and the identification of latent factors that influence performance.

Acquiring accurate causal relationships, however, is far from trivial. Manual modeling is rarely feasible, as even system experts can only approximate the underlying probabilities or distributions, which are often non-linear and highly dynamic \cite{pham2024causal}. The rapid changes in system states further complicate manual approaches, since probabilities vary depending on the exact scenario \cite{wang2024comprehensivesurveyrootcause}. Data-driven approaches, by contrast, learn causal dependencies directly from observed data and offer the advantage of automation. Examples include constraint-based algorithms, score-based learning, and hybrid methods that combine statistical tests with optimization techniques. While powerful, these methods are typically data-hungry, may suffer from imprecision, and often require full retraining when system dynamics change, making them unsuitable for continuous operation \cite{pham2024causal}.

To address these limitations, \textit{multimodal causality mining} has been proposed. By integrating additional information sources, such as system logs, performance metrics, or contextual metadata, into the analysis process, multimodal approaches can uncover richer causal structures and improve robustness in dynamic environments \cite{li2025context}. Furthermore, different learning strategies have been explored to enable continuous operation \cite{weiss2025review}. Reinforcement learning, for instance, allows causal models to adapt over time while incorporating human feedback. Wang et al. \cite{wang2023hrlhf} demonstrate how hierarchical reinforcement learning can reduce the amount of required human input. Instead of exhaustive edge labeling, which would require $O(N^2)$ decisions, their approach uses efficient pairwise comparison queries based on triplets. However, their method is limited to a single data modality and lacks a fully continuous learning loop. This limitation is shared by most current causality mining techniques, which typically rely on either performance metrics, distributed traces, or system logs in isolation rather than combining multiple data sources for validation \cite{zhang2025failure}. As such, the integration of contextual information into causal inference remains a significant challenge. Addressing this gap is the focus of this paper, which proposes methods to leverage expert insights and structured knowledge to enhance causal reasoning in distributed systems.

\begin{figure*}[!tb]
    \centering
    \includegraphics[width=1\linewidth]{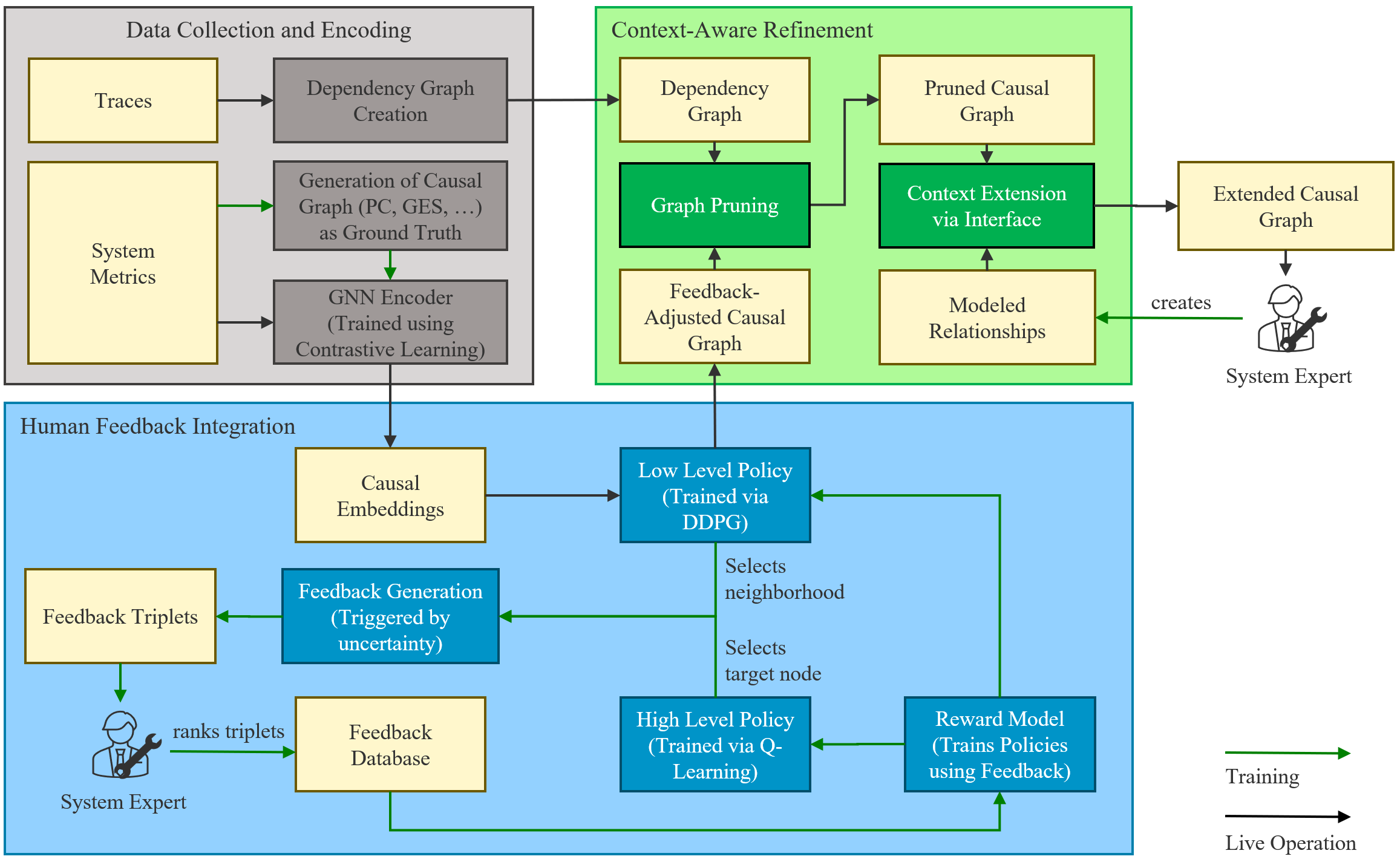}
    \caption{Overview of the continuous analysis pipeline.}
    \label{fig:concept_overview}
\end{figure*}

\section{Concept}\label{sec:concept}
In a previous work, SDVDiag has been introduced as a modular platform for the automated analysis of connected vehicle functions \cite{weiss2025sdvdiag}. This paper extends the existing pipeline by integrating context information into the analysis process. Our concept addresses the limitations of data-driven methods, which have been identified in Section~\ref{sec:causal_inference}, through a multimodal approach. In particular, the following requirements are considered:
\begin{enumerate}
    \item \textbf{Flexible Data Integration}: The pipeline should include multiple metrics simultaneously to capture complex interdependencies between services.
    \item \textbf{Inclusion of Human Feedback}: The approach should incorporate expert knowledge through a structured feedback mechanism that guides the learning process.
    \item \textbf{Integration of System-Specific Information}: The pipeline should use available context such as the system topology to validate and prune statistically inferred edges.
    \item \textbf{Enablement of Additional Context}: The system should provide an extensible interface to refine the graph using different information sources such as application data and contextual rules.
    \item \textbf{Adjustment of the Analysis}: The analysis should maintain its capabilities under high system dynamics through persistent feedback storage and continuous learning.
\end{enumerate}
The resulting approach consists of a reinforcement learning loop that continuously trains the involved model based on expert projections and further refines the causality graph using the system's structure and other modeled relationships. An overview of the analysis pipeline is provided in Fig. \ref{fig:concept_overview}. The process is divided into data collection and encoding, feedback-based generation of a causal graph, and its refinement using context. Further details about these steps are described in the following.




\subsection{Data Collection and Encoding}
SDVDiag relies on established observability tools commonly deployed in cloud-native environments. Prometheus serves as the primary source for performance metrics, collecting time series data that characterize the behavior of individual services and the underlying infrastructure. These metrics include resource utilization indicators such as CPU load and memory consumption, which are scraped at configurable intervals and stored as multivariate time series batches. Additionally, Zipkin provides distributed tracing capabilities, capturing the actual communication flow between system components. These traces document real service invocations and are used to construct dependency graphs that reflect the structural relationships within the monitored system.

The collected multivariate time series data from normal system operation serve as input for a Graph Neural Network (GNN) encoder. This encoder is trained to transform raw metric data into vector representations, referred to as embeddings, that capture causal relationships between services. The fundamental objective is to produce embeddings where causally connected services are mapped to similar vector representations in the embedding space, while services without causal dependencies receive significantly different representations. This property enables the subsequent reinforcement learning system to reason about causal structures based on embedding similarities rather than raw metric values.

The GNN encoder employs contrastive learning for training, where triplets of services are formed: an anchor node, a causally connected positive example, and an unconnected negative example. The training objective pulls positive pairs closer together in the embedding space while pushing negative pairs apart. The encoder follows a modular architecture that supports multiple GNN variants, including Graph Convolutional Networks (GCN) and Graph Attention Networks (GAT), facilitating transfer learning across different system configurations.

The encoder training occurs independently from the rest of the analysis pipeline and takes place at the very beginning of the process. Once pre-trained on representative system data, the encoder remains frozen during all subsequent analysis phases. To enable fully automated training without manual labeling, conventional causal discovery algorithms such as the PC algorithm or Greedy Equivalence Search (GES) are employed to generate initial causal graphs that serve as ground truth for the contrastive learning process. This approach offers the advantage of complete automation, requiring no human intervention during encoder training. However, it also introduces a limitation: since these conventional algorithms are themselves imprecise and may produce erroneous causal structures, the resulting embeddings inherit these inaccuracies. The integration of human feedback aims to address this issue.



\subsection{Human Feedback integration}
The integration of human feedback follows the hierarchical reinforcement learning approach by Wang et al. \cite{wang2023hrlhf}, which SDVDiag extends with a persistent feedback preservation strategy. The core idea is to leverage expert knowledge efficiently through carefully selected comparison queries rather than exhaustive manual annotation.

To reconstruct causal graphs from embeddings, SDVDiag employs a decoder that transforms vector representations back into graph structures. Initially, this decoder uses a simple base policy operating purely on embedding similarity. For any node pair $(x_i, x_j)$, the base policy computes the edge probability as:
\begin{equation}
\pi^l_{base}(x_i, x_j) = \sigma(f(x_i)^\top f(x_j))
\end{equation}
where $\sigma(\cdot)$ denotes the sigmoid function and $f(\cdot)$ the frozen GNN encoder. In practice, this corresponds to cosine similarity between L2-normalized embeddings. An edge is included if this similarity exceeds a threshold $\tau_{base}$. While providing high recall, this similarity-based approach produces many false positives, necessitating refinement through human feedback.

The feedback mechanism uses a hierarchical architecture to minimize expert effort. The system generates triplets consisting of a target node $x_t$ and two candidate predecessors $x_j$ and $x_k$. Experts answer a binary comparison: which candidate is more likely a causal predecessor? This pairwise format scales more gracefully than exhaustive edge labeling, which would require $O(N^2)$ decisions.

The low-level policy $\pi_l$ serves dual purposes: during operation, it generates causal edges and their probabilities, replacing the base policy after training; during feedback generation, it identifies edges with highest uncertainty for a given target node. Training uses Deep Deterministic Policy Gradient (DDPG), an actor-critic algorithm for continuous action spaces that outputs edge probabilities in $[0, 1]$ and employs experience replay to efficiently reuse expensive feedback samples.

The high-level policy $\pi_h$ operates at a strategic level, selecting the most informative target nodes for detailed optimization. It employs Q-learning with $\epsilon$-greedy exploration, well-suited for its discrete binary action space (select or skip). The optimal decision follows from direct Q-value comparison, while $\epsilon$-greedy ensures sufficient exploration with lower complexity than more sophisticated alternatives.

Feedback collection operates as a continuous background process. Triplets are generated asynchronously, and retraining is triggered once pending queries exceed a threshold. This batch approach stabilizes the system by preventing individual responses from causing abrupt structural changes. All feedback is stored in a persistent database maintaining the complete history of expert preferences as triplets $(x_t, x_{preferred}, x_{rejected})$, enabling incremental learning and preserving knowledge across system restarts.

The collected feedback trains a reward model $r_\theta$ that approximates expert preferences. This neural network receives concatenated embeddings of target and candidate nodes, producing a quality score for the potential causal relationship. Training employs cross-entropy loss for pairwise rankings. The reward model provides training signals for both policies with distinct aggregation levels:
\begin{equation}
r_{low}(x_t, x_j) = r_\theta(x_t, x_j)
\end{equation}
\begin{equation}
r_{high}(x_t) = \sum_{x_j \in \mathcal{N}(x_t)} r_\theta(x_t, x_j)
\end{equation}
Due to this design, the low-level policy aims to optimize individual edges directly, while the high-level policy learns to prioritize nodes with many promising causal relationships through aggregated neighborhood rewards.


\begin{figure*}[!t]
\centering
\subfloat[Parking lot of the UGV fleet.]{\includegraphics[width=2.5in]{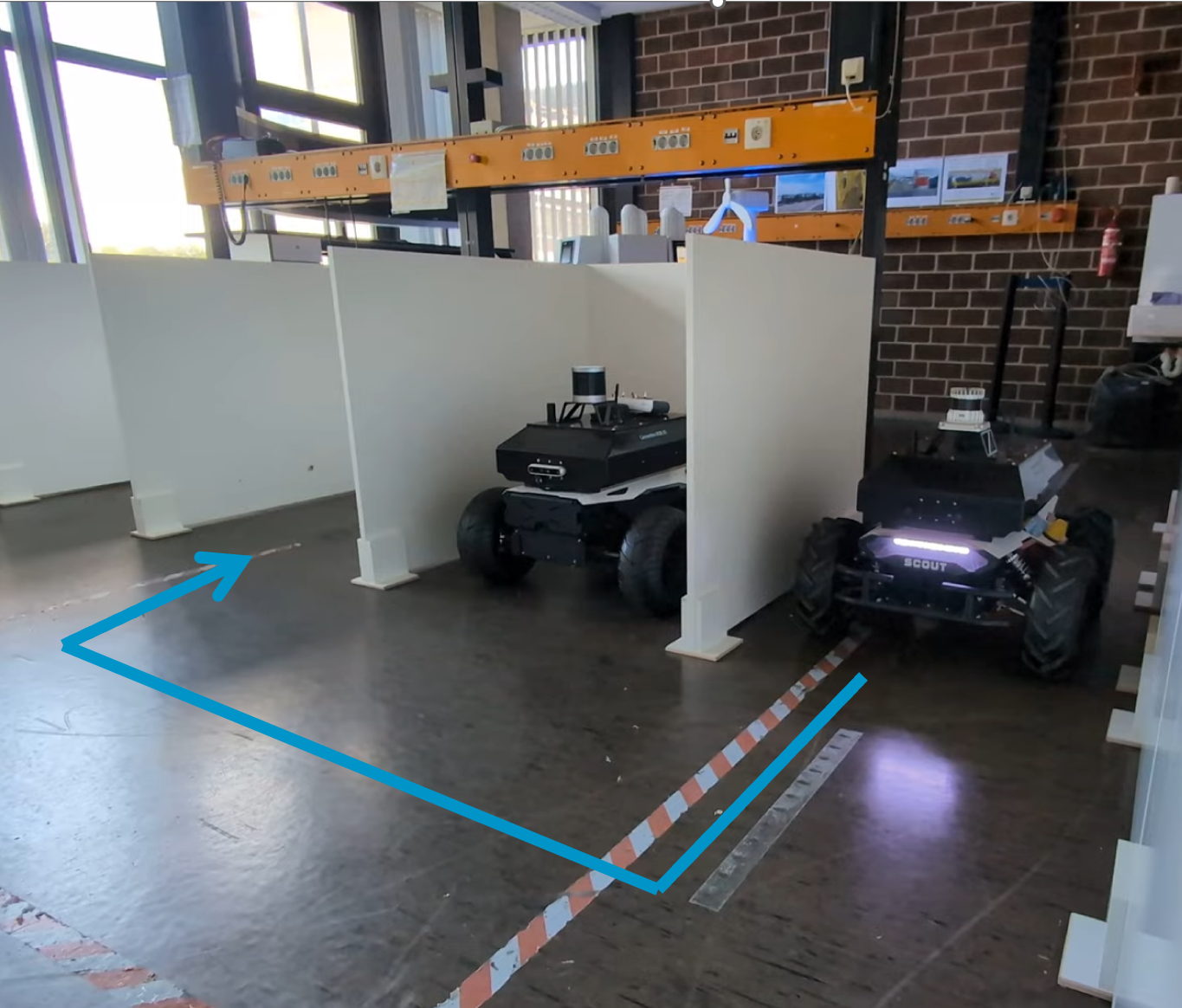}%
\label{fig:valet_parking_overview}}
\hfil
\subfloat[Expert-estimated causal graph of the application.]{\includegraphics[width=2.5in]{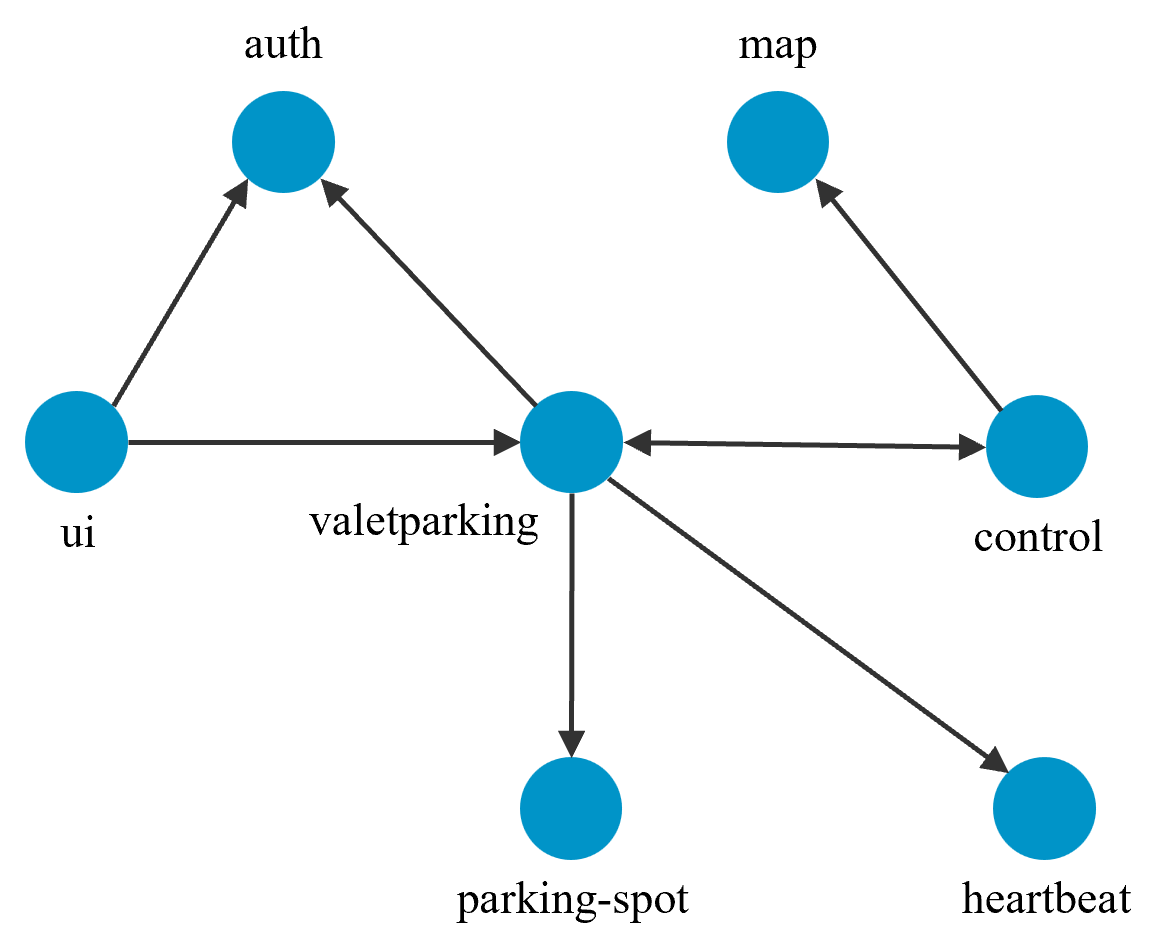}%
\label{fig:valet_parking_services}}
\caption{Automated Valet Parking application overview.}
\label{fig:valet_parking}
\end{figure*}


\subsection{Context-Aware Refinement}
Before the feedback-adjusted causal graph is presented to system engineers, its edges undergo additional refinement based on provided context. Even after human feedback integration, spurious correlations may persist due to statistical artifacts in the metric data. To address this, SDVDiag employs a multimodal pruning mechanism that validates learned edges against distributed tracing data from Zipkin, which records actual service invocations and provides ground truth for communication relationships.

The pruning process applies three sequential filters. First, confidence filtering removes edges whose prediction probability falls below a configurable threshold, eliminating low-confidence predictions. Second, trace validation cross-references each edge against the Zipkin dependency graph to verify that a corresponding service-level communication exists. Edges between metrics whose parent services show no documented interaction are discarded as structurally implausible. Third, duplicate aggregation consolidates multiple metric-level edges between the same service pair, retaining only the variant with highest confidence.

A key challenge in trace validation is the abstraction level mismatch between data sources. The multimodal graph operates at pod-metric granularity, where nodes represent specific metrics of individual Kubernetes pods. In contrast, Zipkin captures dependencies at service level, independent of specific metrics or pod instances. To bridge this gap, service names are extracted from metric labels by removing metric suffixes and Kubernetes deployment identifiers, enabling validation of metric-level edges against service-level communication patterns. Since metric correlations can be bidirectional while Zipkin captures actual call direction, the filter also corrects edge orientation when the reverse direction exists in the trace data.

After generating the pruned graph, it is subsequently enriched through a context extension module that enables integration of expert-defined causal relationships. System experts can specify additional nodes and edges in a structured format, defining rules that trigger under specific conditions. These rules can reference existing graph elements or current system metrics. For example, a rule might specify: if CPU utilization of a service exceeds a defined threshold, inject a node representing an external component and add an edge connecting it to the affected metric. This mechanism captures causal factors that are difficult to observe directly, such as message queue backlogs that influence downstream services without appearing in standard monitoring. In addition, the interface enables the automated extraction and transformation of causal relationships from other system models or application data.

To maintain graph validity, each proposed relationship undergoes cycle detection before integration. Since causal graphs must form directed acyclic graphs (DAGs), any edge that would introduce a cycle is rejected. The resulting graph combines statistical inference with structural validation and additional context information, providing a robust foundation for subsequent root cause analysis.

\section{Evaluation}\label{sec:evaluation}

The proposed approach has been implemented as an extension to the SDVDiag platform and deployed alongside an Automated Valet Parking (AVP) application. The AVP system coordinates a fleet of Unmanned Ground Vehicles (UGVs) that autonomously perform parking operations, communicating with backend services for authentication, map provision, parking spot management, and vehicle control. Fig.~\ref{fig:valet_parking_services} shows the service dependency graph of the AVP application, which serves as ground truth for evaluating the generated causal graphs. The causal edges in this reference graph have been estimated by domain experts based on their understanding of the system architecture and communication patterns.

Both the AVP application and the analysis platform operate in live mode on a Kubernetes cluster comprising one control node and five worker nodes, with one node dedicated to the analysis platform. The control node hosts the observability stack including Prometheus for metric collection and Zipkin for distributed tracing, while the worker nodes execute the AVP services as containerized pods.

The evaluation scenario simulates a queue overflow condition caused by an overload of parking requests. This leads to synchronization issues between the parking lot management and the parking spot manager service. From an external perspective (see Fig.~\ref{fig:valet_parking_overview}), the resulting behavior manifests as a UGV that stops at the drop-off location and remains stationary despite free parking spots being available. The following subsections use this scenario to verify the functionality of the proposed approach through its various processing stages.

\begin{figure*}[!t]
\centering
\subfloat[Raw causal graph with false positives.]{\includegraphics[width=3.5in]{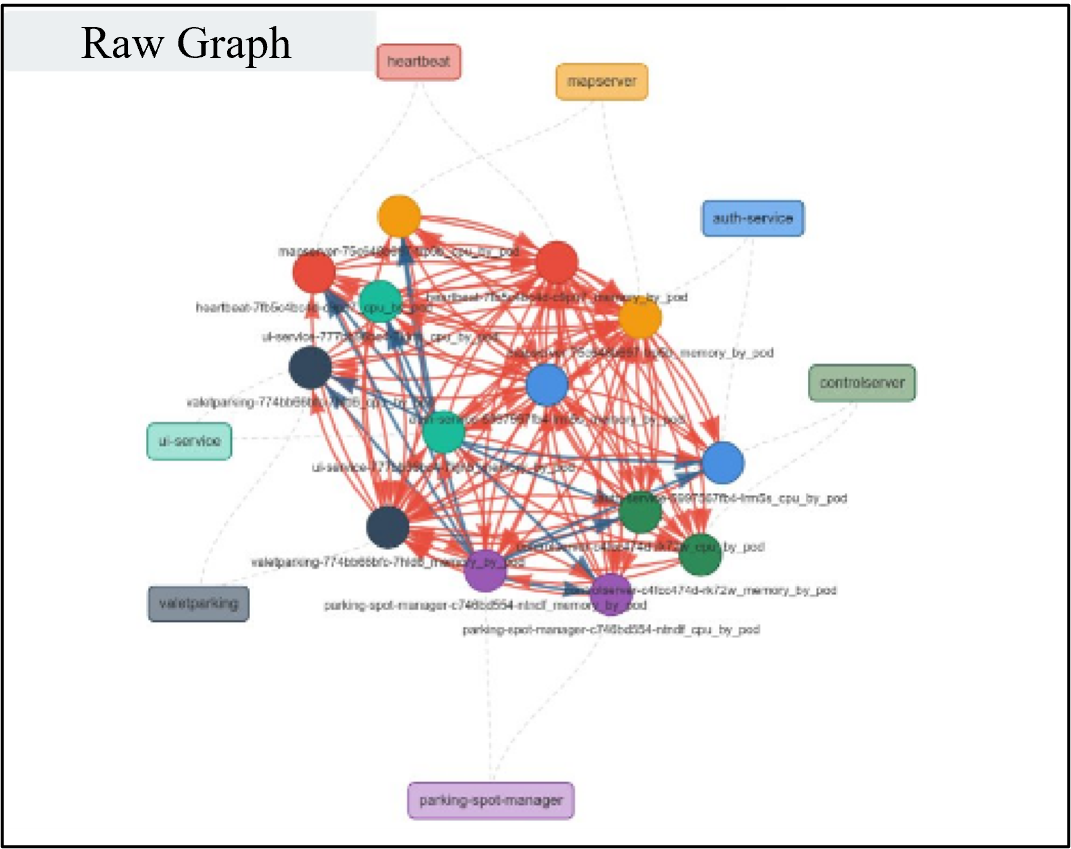}%
\label{fig:raw_graph}}
\hfil
\subfloat[Context-refined causal graph. Green edges match the ground truth, red edges are missing.]{\includegraphics[width=3.5in]{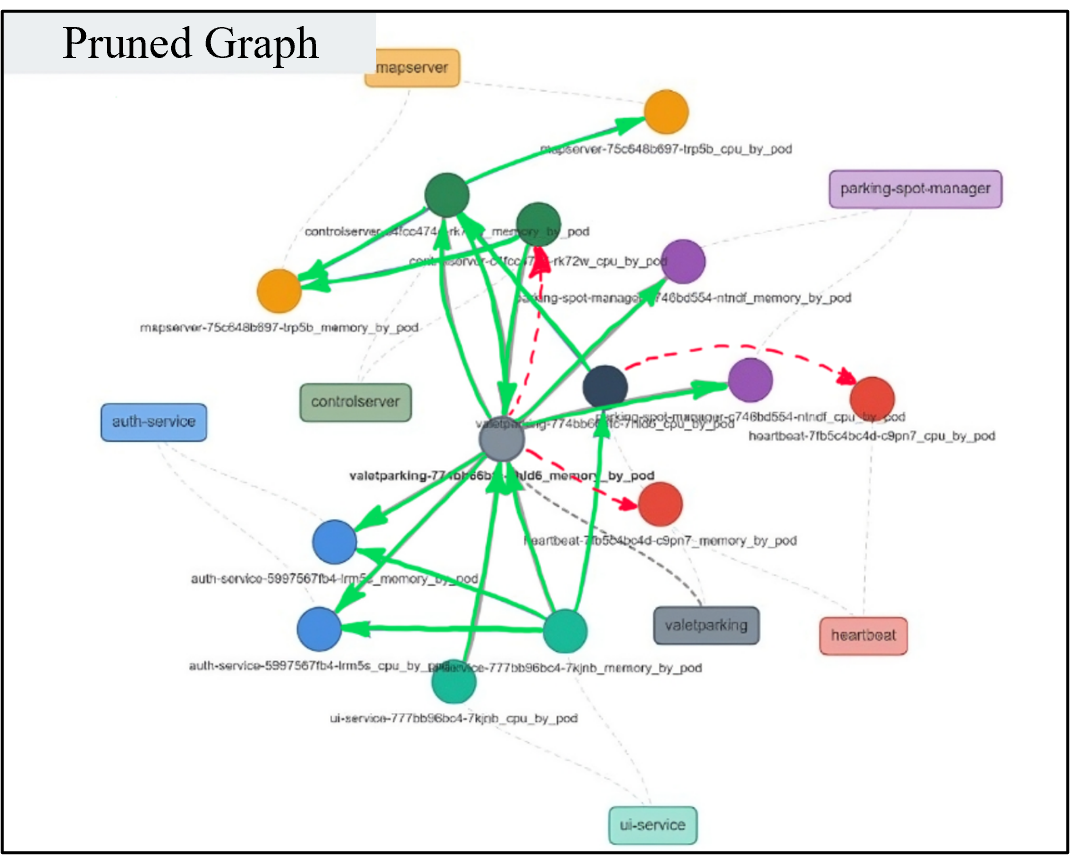}%
\label{fig:context_graph}}
\caption{Causal Graph (a) before and (b) after the refinement via context.}
\label{fig:graph_evolution}
\end{figure*}


\subsection{Causal Graph Refinement}

Fig.~\ref{fig:graph_evolution} illustrates the causal graph before and after processing by the analysis pipeline. A comparison between all processing stages and the ground truth can be found in Table~\ref{tab:graph_comparison}. 

Fig.~\ref{fig:raw_graph} shows the raw causal graph generated by the base policy. On average across all experiments, this graph contains around 137 edges with many false positives due to the limitations of purely statistical inference on observational data.

The intermediate, feedback-adjusted graph is created after collection of 30 expert responses. The trained reward model and optimized RL policies eliminate implausible dependencies that experts identified during the feedback process. On average, around 49 edges remain in the graph, with central services emerging as hubs with multiple well-defined dependencies.

Fig.~\ref{fig:context_graph} shows the pruned graph after trace validation. The three-stage pruning pipeline reduces the graph to 16 edges by applying confidence filtering, Zipkin validation, and duplicate aggregation. The pruned graph achieves 100\% precision when compared with the ground truth, meaning all identified edges correspond to causal relationships. The 84\% recall indicates that 16 of 19 expected causal edges were correctly identified. The missing edges primarily concern services with low invocation frequency during the test period.

\subsection{Context-Aware Analysis}

The final evaluation step demonstrates the context extension module by injecting context information about the valet parking application. In the evaluation scenario, the parking queue overflow represents a causal factor that does not appear in standard monitoring since it is not implemented as a separate service with its own metrics. To capture this hidden dependency, the rule shown in Listing~\ref{lst:parking_rule} is added.

\begin{lstlisting}[language=yaml,caption={Context rule for parking queue overflow detection.},label={lst:parking_rule}]
rule_id: "parking_queue_overflow"
condition:
  metric: "valetparking_cpu_by_pod"
  threshold: 80.0
  operator: ">"
inject_node: "parking_queue"
inject_edge:
  from: "parking_queue"
  to: "valetparking_cpu_by_pod"
\end{lstlisting}

\begin{figure}[!tb]
    \centering
    \includegraphics[width=1\linewidth]{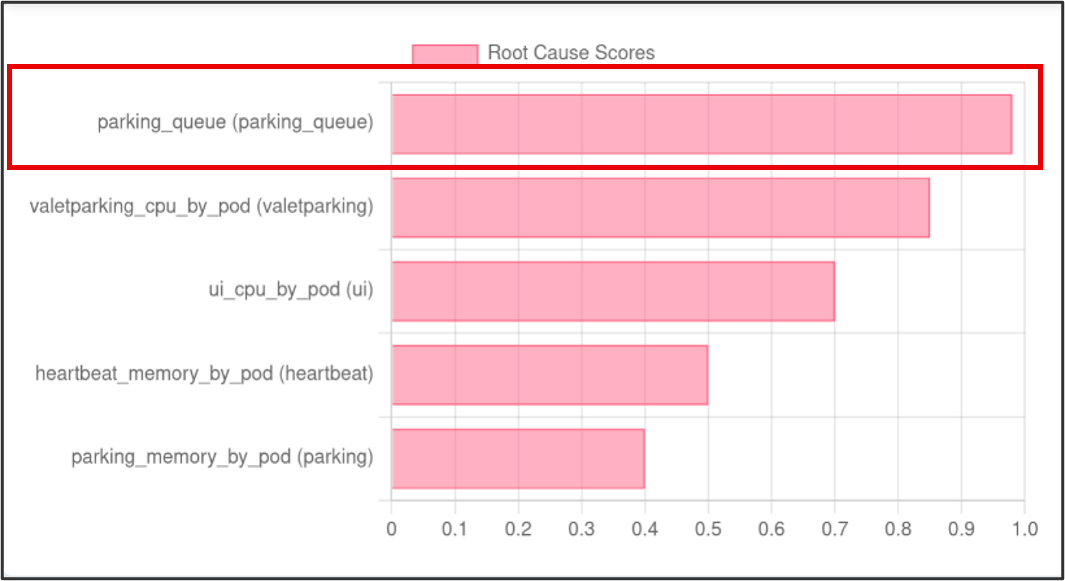}
    \caption{Resulting root-cause analysis for the given error scenario.}
    \label{fig:rca}
\end{figure}

When the valetparking service CPU utilization exceeds the specified threshold during the queue overflow scenario, the rule activates and injects a \texttt{parking\_queue} node with a causal edge to the affected metric. As shown in Fig.~\ref{fig:rca}, this extension enables the root cause analysis (performed via random walk on the final graph) to correctly identify the queue overflow as the actual cause of the observed system behavior, rather than attributing the fault to downstream effects such as increased CPU load or synchronization failures.

Furthermore, the context extension module thus demonstrates its potential for hypothesis testing and integration of expert knowledge about hidden system components. System experts can formulate causal hypotheses as rules and observe whether the extended graph improves diagnostic accuracy. This capability is particularly valuable in complex distributed systems where not all causal factors are directly observable through standard telemetry.

\begin{table}[!b]
\centering
\caption{Comparison of graph versions against ground truth.}
\label{tab:graph_comparison}
\begin{tabular}{lcccc}
\hline
\textbf{Graph Version} & \textbf{Edges} & \textbf{Precision} & \textbf{Recall} & \textbf{F1} \\
\hline
Ground Truth (expected) & 19 & -- & -- & -- \\
Raw (Base Policy) & 137 & 14\% & 100\% & 24\%  \\
Feedback-adjusted & 49 & 32\% & 84\% & 47\% \\
Pruned (final) & 16 & 100\% & 84\% & 91\% \\
\hline
\end{tabular}
\end{table}

\section{Conclusion}\label{sec:conclusion}

This paper presented a context-aware approach for causality mining in connected vehicle systems that addresses key limitations of purely data-driven methods. By extending the SDVDiag platform with hierarchical reinforcement learning from human feedback, multimodal graph pruning, and a context extension interface, the approach enables the integration of expert knowledge and system-specific information into the causal discovery process.

The evaluation on an Automated Valet Parking application demonstrated significant improvements in diagnostic precision. The multimodal pruning pipeline reduced the initial graph from 137 edges to 16 validated edges while achieving up to 100\% precision and 84\% recall against the expert-defined ground truth. The complementary effect of human feedback and trace-based validation proved particularly effective: while the learning procedure captures complex dependency patterns from multivariate time series, the pruning module systematically eliminates false positives through structural validation against actual service communication. Furthermore, the context extension module successfully identified hidden causal factors such as queue overflows that do not appear in standard telemetry, enabling more accurate root cause analysis.

Despite these promising results, several limitations must be acknowledged. First, the evaluation was conducted on a relatively small system with seven services, and scalability to larger microservice architectures with hundreds of components remains to be validated. Second, the ground truth used for evaluation relies on expert projections rather than definitive causal knowledge, which represents a general limitation in the field of causal discovery. Third, the available data regarding error scenarios was limited, as an accurate simulation of failure patterns through error injection remains difficult in purely experimental environments.

Future work should address these limitations through several directions. Large-scale experiments on more complex microservice systems would provide stronger evidence for the approach's practical applicability. The GNN encoder and causal embedding generation offer substantial room for improvement, as embedding quality directly impacts downstream policy performance. Alternative contrastive learning strategies and more sophisticated graph neural network architectures could yield more discriminative representations. Additionally, further techniques for combining explicit model information with data-driven learning should be explored, potentially including the integration of logs and system documentation for improved interpretability.

\section*{Acknowledgment}
This work has been funded by the German Federal Ministry of Research, Technology and Space (BMFTR) under grant 02P23Q843.

The authors disclose that generative AI has been used to improve grammar and language of this publication. The authors have reviewed and edited all content as needed and take full responsibility for its scientific integrity and authenticity.

\bibliographystyle{IEEEtran}
\bibliography{bib}

\end{document}